\documentclass[12pt]{spieman} 
\usepackage{amsmath} 
\usepackage[]{graphicx}
\usepackage{setspace}
\usepackage{tocloft}
\usepackage{multirow}

\newcommand{\refnum}[1]{Ref.~\citenum{#1}}
\newcommand{\reffig}[1]{Fig.~\ref{#1}}
\newcommand{\reftable}[1]{Table \ref{#1}} 

\title{WFIRST-AFTA Coronagraph Science Yield Modeling with EXOSIMS} 

\author{Dmitry Savransky\supscr{a,b} and Daniel Garrett\supscr{a}}

\affiliation{\supscrsm{a}Sibley School of Mechanical and Aerospace Engineering, Cornell University, Ithaca, NY, USA 14853\\
\supscrsm{b}Carl Sagan Institute, Cornell University, Ithaca, NY, USA 14853}

\cftpagenumbersoff{figure}
\cftpagenumbersoff{table} 

\begin{document} 
\maketitle 

\begin{abstract}
We present and discuss the design details of an extensible, modular, open source software framework called EXOSIMS, which creates end-to-end simulations of space-based exoplanet imaging missions. We motivate the development and baseline implementation of the component parts of this software with models of the WFIRST-AFTA coronagraph, and present initial results of mission simulations for various iterations of the WFIRST-AFTA coronagraph design.  We present and discuss two sets of simulations: The first compares the science yield of completely different instruments in the form of early competing coronagraph designs for WFIRST-AFTA. The second set of simulations evaluates the effects of different operating assumptions, specifically the assumed post-processing capabilities and telescope vibration levels.  We discuss how these results can guide further instrument development and the expected evolution of science yields.
\end{abstract}


{\noindent \footnotesize{\bf Address all correspondence to}: Dmitry Savransky (\linkable{ds264@cornell.edu})}

\begin{spacing}{2}   

\section{Introduction}
\label{sect:intro}  
The majority of exoplanets discovered to date have been detected indirectly, by looking for effects these planets have on their host stars. Directly imaging exoplanets will provide a great deal of additional information unobtainable by most indirect detection methods and make discoveries expanding the population of known exoplanets. While direct imaging of exoplanets has been demonstrated with ground based instruments, these have all been very young, very large, and self-luminous planets on long-period orbits. Imaging of smaller and more Earth-like planets will likely require space observatories such as the Wide-Field Infrared Survey Telescope-Astrophysics Focused Telescope Assets (WFIRST-AFTA). Such observatories are major undertakings requiring extensive planning and design.

Building confidence in a mission concept's ability to achieve its science goals is always desirable.  Unfortunately, accurately modeling the science yield of an exoplanet imager can be almost as complicated as designing the mission.  While each component of the system is modeled in great detail as it proceeds through its design iterations, fitting these models together is very challenging. Making statements about expected science returns over the course of the whole mission requires a large number of often unstated assumptions when such results are presented.  This makes it challenging to compare science simulation results and also to systematically test the effects of changing just one part of the mission or instrument design from different groups.

We seek to address this problem with the introduction of a new modular, open source mission simulation tool called EXOSIMS (Exoplanet Open-Source Imaging Mission Simulator).  This software is specifically designed to allow for systematic exploration of exoplanet imaging mission science yields. The software framework makes it simple to change the modeling of just one aspect of the instrument, observatory, or overall mission design.  At the same time, this framework allows for rapid prototyping of completely new mission concepts by reusing pieces of previously implemented models from other mission simulations.


Modeling the science yield of an exoplanet imager is primarily difficult because it is completely conditional on the true distributions of planet orbital and physical parameters, of which we so far have only partial estimates. This makes the mission model an inherently probabilistic one, which reports posterior distributions of outcomes conditioned on some selected priors.  Since the introduction of observational completeness by Robert Brown\cite{brown2005}, it is common to approach exoplanet mission modeling with Monte Carlo methods.  Various groups have pursued such modeling, often focusing on specific aspects of the overall mission or observation modeling\cite{brown2010new,Savransky2010,turnbull2012search,Stark2014}.

A second challenge is correctly including all of the dynamic and stochastic aspects of such a mission. Given a spacecraft orbit, a target list, and the constraints of the imaging instrument, we can always predict when targets will be observable.  Incorporating this knowledge into a simulation, however, can be challenging if a single calculated value represents the predictions, i.e., the number of planets discovered.  Similarly, while it is simple to write down the probability of detecting a planet upon the first observation of a star, it is more challenging to do the same for a second observation an arbitrary amount of time later, without resorting to numerical simulation\cite{brown2010new}.  EXOSIMS deals with these challenges by explicitly simulating every aspect of the mission and producing a complete timeline of simulated observations including the specific targets observed at specific times in the mission and recording the simulated outcomes of these observations.  While one such simulation does not answer the question of expected mission science yield, an ensemble of many thousands of such simulations gives the data for the posterior distributions of science yield metrics. EXOSIMS is designed to generate these ensembles and provide the tools to analyze them, while allowing the user to model any aspect of the mission as detailed as desired.

In \S\ref{sec:EXOSIMS} we provide an overview of the software framework and some details on its component parts.  As the software is intended to be highly reconfigurable, we focus on the operational aspects of the code rather than implementation details. We use the coronagraphic instrument currently being developed for WFIRST-AFTA as a motivating example for specific implementations of the code.  In \S\ref{sec:wfirst} we present mission simulation results for various iterations of the WFIRST-AFTA coronagraph designs using components that are being adapted to build the final implementation of EXOSIMS.

EXOSIMS is currently being developed as part of a WFIRST Preparatory Science investigation, with initial implementation targeted at WFIRST-AFTA.  This development includes the definition of a strict interface control, along with corresponding prototypes and class definitions for each of the modules described below.  The interface control document and as-built documentation are both available for public review and comment at \linkable{https://github.com/dsavransky/EXOSIMS}.  Initial code release is targeted for Fall 2015, with an alpha release in February of 2016 and continued updates through 2017. 

Future development of EXOSIMS is intended to be a community-driven project, and all software related to the base module definitions and simulation execution will be made publicly available alongside the interface control documentation to allow mission planners and instrument designers to quickly write their own modules and drop them directly into the code without additional modifications made elsewhere. We fully expect that EXOSIMS will be highly useful for ensuring the achievement of the WFIRST-AFTA science goals, and will be of use to the design and planning of future exoplanet imaging missions.

\section{EXOSIMS Description}\label{sec:EXOSIMS}
EXOSIMS builds upon previous frameworks described in \refnum{Savransky2010} and \refnum{Savransky2013}, but will be significantly more flexible than these earlier efforts, allowing for seamless integration of independent software modules, each of which performs its own well-defined tasks, into a unified mission simulation.  This will allow the wider exoplanet community to quickly test the effects of changing a single set of assumptions (for example, the specific model of planet spectra, or a set of mission operating rules) on the overall science yield of the mission, by only updating one part of the simulation code rather than rewriting the entire simulation framework.  

The terminology used to describe the software implementation is loosely based on the object-oriented framework upon which EXOSIMS is built.  The term module can refer to either the object class prototype representing the abstracted functionality of one piece of the software, or to an implementation of this object class which inherits the attributes of the prototype, or to an instance of this object class.  Thus, when we speak of input/output definitions of modules, we are referring to the class prototype.  When we discuss implemented modules, we mean the inherited class definition.  Finally, when we speak of passing modules (or their outputs), we mean the instantiation of the inherited object class being used in a given simulation.  Relying on strict inheritance for all implemented module classes provides an automated error and consistency-checking mechanism, as we can always compare the outputs of a given object instance to the outputs of the prototype.  This means that it is trivial to pre-check whether a given module implementation will work with the larger framework, and thus allows for the flexibility and adaptability described above. 

\begin{figure}[ht]
    \centering
        \includegraphics[width=\textwidth]{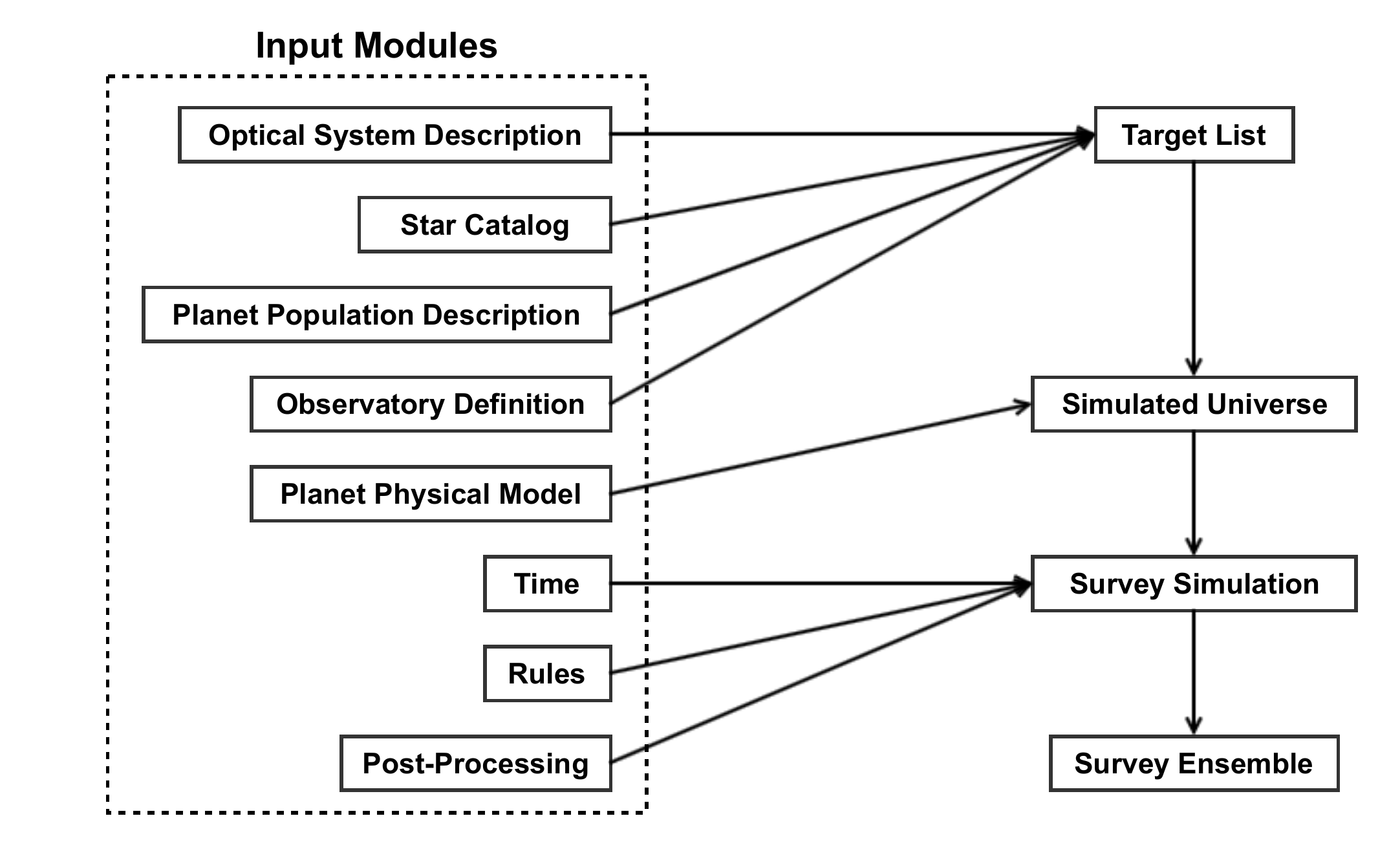}
   \caption{\label{fig:codeflow} Flowchart of mission simulation. Each box represents a component software module which interacts with other modules as indicated by the arrows. The simulation modules (those that are not classified as input modules) pass all input modules along with their own output.  Thus, the Survey Ensemble module has access to all of the input modules and all of the upstream simulation modules.} 
\end{figure} 

\reffig{fig:codeflow} shows the relationships of the component software modules classified as either input modules or simulation modules. The input modules contain specific mission design parameters. The simulation modules take the information contained in the input modules and perform mission simulation tasks.  Any module may perform any number or kind of calculations using any or all of the input parameters provided.  They are only constrained by their input and output specification, which is designed to be as flexible as possible, while limiting unnecessary data passing to speed up execution.

\subsection{Input Modules}

The specific mission design under investigation determines the functionality of each of the input modules, but the inputs and outputs of each are always the same (in terms of data type and what the variables represent). These modules encode and/or generate all of the information necessary to perform mission simulations.  Here we briefly describe the functionality and major tasks for each of the input modules.

\subsubsection{Optical System Description}
The Optical System Description module contains all of the necessary information to describe the effects of the telescope and starlight suppression system on the target star and planet wavefronts.  This requires encoding the design of both the telescope optics and the specific starlight suppression system, whether it be an internal coronagraph or an external occulter.  The encoding can be achieved by specifying Point Spread Functions (PSF) for on- and off-axis sources, along with (potentially angular separation-dependent) contrast and throughput definitions.  At the opposite level of complexity, the encoded portions of this module may be a description of all of the optical elements between the telescope aperture and the imaging detector, along with a method of propagating an input wavefront to the final image plane.  Intermediate implementations can include partial propagations, or collections of static PSFs representing the contributions of various system elements.  The encoding of the optical train will allow for the extraction of specific bulk parameters including the instrument inner working angle (IWA), outer working angle (OWA), and mean and max contrast and throughput.

If the starlight suppression system includes active wavefront control, i.e., via one or more deformable mirrors (DM) \cite{cahoy2014wavefront}, then this module must also encode information about the sensing and control mechanisms.  Again, this can be achieved by simply encoding a static targeted DM shape, or by dynamically calculating DM settings for specific targets via simulated phase retrieval.  As wavefront control residuals may be a significant source of error in the final contrast budget, it is vitally important to include the effects of this part of the optical train.

The optical system description can optionally include stochastic and systematic wavefront-error generating components.  Again, there is a wide range of possible encodings and complexities.  They could be Gaussian errors on the contrast curves sampled during survey simulation to add a random element to the achieved contrast on each target.  Alternatively, in cases where an active wavefront control system is modeled, stochastic wavefront errors could be introduced by simulating the measurement noise on the wavefront sensor (either again as drawn from pre-determined distributions, or additively from various detector and astrophysical noise sources).  Systematic errors, such as mis-calibration of deformable mirrors, closed-loop control delays, and non-common path errors, may be included to investigate their effects on contrast or optical system overhead.  In cases where the optical system is represented by collections of static PSFs, these effects must be included in the diffractive modeling that takes place before executing the simulation.  For external occulters, we draw on the large body of work on the effects of occulter shape and positioning errors on the achieved contrast, as in \refnum{shaklan2010error}.

Finally, the optical system description must also include a description of the science instrument or instruments.  The baseline instrument is assumed to be an imaging spectrometer, but pure imagers and spectrometers are also supported.  Each instrument encoding must provide its spatial and wavelength coverage and sampling. Detector details such as read noise, dark current, and quantum efficiency must be provided, along with more specific quantities such as clock induced charge for electron multiplying CCDs\cite{denvir2003electron}. Optionally, this portion of the module may include descriptions of specific readout modes, i.e., in cases where Fowler sampling\cite{fowler1990demonstration} or other noise-reducing techniques are employed. In cases where multiple science instruments are defined, they are given enumerated indices in the specification, and the Survey Simulation module must be implemented so that a particular instrument index is used for a specific task, i.e., detection vs. characterization. 

The overhead time of the optical system must also be provided and is split into two parameters.  The first is an integration time multiplier for detection and characterization modes, which represents the individual number of exposures that need to be taken to cover the full field of view, full spectral band, and all polarization states in cases where the instrument splits polarizations.  For detection modes, we will typically wish to cover the full field of view, while possibly only covering a small bandpass and only one polarization, whereas for characterizations, we will typically want all polarizations and spectral bands, while focusing on only one part of the field of view.  The second overhead parameter gives a value for how long it will take to reach the instrument's designed contrast on a given target.  This overhead is separate from the one specified in the observatory definition, which represents the observatory settling time and may be a function of orbital position, whereas the contrast floor overhead may depend on target brightness.  If this value is constant, as in the case of an observing strategy where a bright target is used to generate the high contrast regions, or zero, as in the case of an occulter, then it can be folded in with the observatory overhead.

\subsubsection{Star Catalog}
The Star Catalog module includes detailed information about potential target stars drawn from general databases such as SIMBAD\cite{wenger2000simbad}, mission catalogs such as Hipparcos\cite{perryman1997hipparcos}, or from existing curated lists specifically designed for exoplanet imaging missions\cite{turnbull2012search}.  Information to be stored, or accessed by this module will include  target positions and proper motions at the reference epoch (see \S\ref{sec:time}), catalog identifiers (for later cross-referencing), bolometric luminosities, stellar masses, and magnitudes in standard observing bands.  Where direct measurements of any value are not available, values are synthesized from ancillary data and empirical relationships, such as color relationships and mass-luminosity relations\cite{henry2004}.

This module will not provide any functionality for picking the specific targets to be observed in any one simulation, nor even for culling targets from the input lists where no observations of a planet could take place.  This is done in the Target List module as it requires interactions with the Planetary Population module (to determine the population of interest), the Optical System Description module (to define the capabilities of the instrument), and Observatory Definition module (to determine if the view of the target is unobstructed).

\subsubsection{Planet Population Description}
The Planet Population Description module encodes the density functions of all required planetary parameters, both physical and orbital. These include semi-major axis, eccentricity, orbital orientation, and planetary radius and mass. Certain parameter models may be empirically derived\cite{Savransky2011} while others may come from analyses\cite{Dressing2013,Fortney2007} of observational surveys such as the Keck Planet Search\cite{Cumming2008,Howard2010}, Kepler\cite{Batalha2013,Fressin2013,Petigura2013}, and ground-based imaging surveys including the Gemini Planet Imager Exoplanet Survey\cite{McBride2011,Macintosh2014}.  This module also encodes the limits on all parameters to be used for sampling the distributions and determining derived cutoff values such as the maximum target distance for a given instrument's IWA.

The Planet Population Description module does not model the physics of planetary orbits or the amount of light reflected or emitted by a given planet, but rather only encodes the statistics of planetary occurrence and properties.  As this encoding is based on density functions, it fully supports modeling `toy' universes where all parameters are fixed, in which case all of the distributions become delta functions.  We can equally use this encoding to generate simulated universes containing only `Earth-twins' to compare with previous studies as in \refnum{brown2005} or \refnum{Stark2014}.  Alternatively, the distributions can be selected to mirror, as closely as possible, the known distributions of planetary parameters.  As this knowledge is limited to specific orbital or mass/radius scales, this process invariably involves some extrapolation.

\subsubsection{Observatory Description}
\begin{figure}[ht]
\centering
\includegraphics[width=1\textwidth]{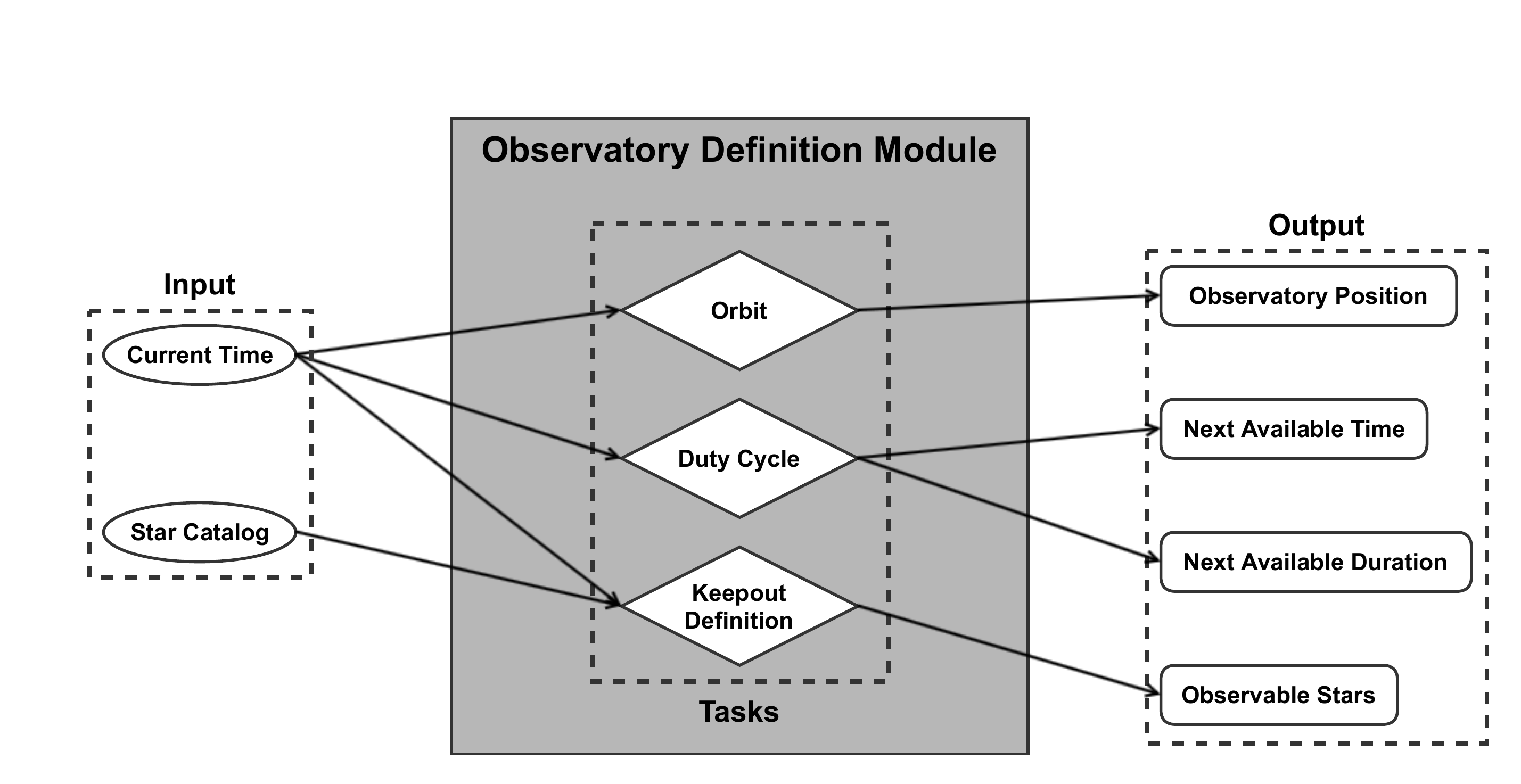}
\caption{ \label{fig:observatory} Depiction of Observatory Definition module including inputs, tasks, and outputs.} 
\end{figure}

The Observatory Definition module contains all of the information specific to the space-based observatory not included in the Optical System Description module. The module has three main tasks: orbit, duty cycle, and keepout definition, which are implemented as functions within the module. The inputs and outputs for these functions are represented schematically in \reffig{fig:observatory}.

The observatory orbit plays a key role in determining which of the target stars may be observed for planet finding at a specific time during the mission lifetime. The Observatory Definition module's orbit function takes the current mission time as input and outputs the observatory's position vector. The position vector is standardized throughout the modules to be referenced to a heliocentric equatorial frame at the J2000 epoch. The observatory's position vector is used in the keepout definition task and Target List module to determine which of the stars from the Star Catalog may be targeted for observation at the current mission time.

The duty cycle determines when during the mission timeline the observatory is allowed to perform planet-finding operations. The duty cycle function takes the current mission time as input and outputs the next available time when exoplanet observations may begin or resume, along with the duration of the observational period. The outputs of this task are used in the Survey Simulation module to determine when and how long exoplanet finding and characterization observations occur.  The specific implementation of the duty cycle function can have significant effects on the science yield of the mission.  For example, if the observing program is pre-determined, such that exoplanet observations can only occur at specific times and last for specific durations, this significantly limits the observatory's ability to respond dynamically to simulated events, such as the discovery of an exoplanet candidate.  This can potentially represent a sub-optimal utilization of mission time, as it may prove to be more efficient to immediately spectrally characterize good planetary candidates rather than attempting to re-observe them at a later epoch.  It also limits the degree to which followup observations can be scheduled to match the predicted orbit of the planet.  Alternatively, the duty cycle function can be implemented to give the exoplanet observations the highest priority, such that all observations can be scheduled to attempt to maximize dynamic completeness\cite{brown2010new} or some other metric of interest. 

The keepout definition determines which target stars are observable at a specific time during the mission simulation and which are unobservable due to bright objects within the field of view such as the sun, moon, and solar system planets.  The keepout volume is determined by the specific design of the observatory and, in certain cases, by the starlight suppression system.  For example, in the case of external occulters, the sun cannot be within the 180$^\circ$ annulus immediately behind the telescope (with respect to the line of sight) as it would be reflected by the starshade into the telescope.  The  keepout definition function takes the current mission time and Star Catalog module output as inputs and outputs a list of the target stars which are observable at the current time. It constructs position vectors of the target stars and bright objects which may interfere with observations with respect to the observatory. These position vectors are used to determine if bright objects are in the field of view for each of the potential stars under exoplanet finding observation. If there are no bright objects obstructing the view of the target star, it becomes a candidate for observation in the Survey Simulation module.

The observatory definition also includes the target transition time, which encodes the amount of overhead associated with transitioning to a new target before the next observation can begin.  For missions with external occulters, this time includes both the transit time between targets as well as the time required to perform the fine alignment at the end of the transit.  For internal coronagraphs, this includes the settling time of the telescope to reach the bus stability levels required by the active wavefront control system.  These may all be functions of the orbital position of the telescope, and may be implemented to take into account thermal effects when considering observatories on geocentric orbits.  This overhead calculation does not include any additional time required to reach the instrument's contrast floor, which may be a function of target brightness, and is encoded separately in the Optical System Description.

In addition to these functions, the observatory definition can also encode finite resources that are used by the observatory throughout the mission.  The most important of these is the fuel used for stationkeeping and repointing, especially in the case of occulters which must move significant distances between observations.  We could also consider the use of other volatiles such as cryogens for cooled instruments, which tend to deplete solely as a function of mission time.  This module also allows for detailed investigations of the effects of orbital design on the science yield, e.g., comparing the baseline geosynchronous 28.5$^\circ$ inclined orbit for WFIRST-AFTA\cite{Spergel2013} with an alternative L2 halo orbit also proposed for other exoplanet imaging mission concepts \cite{savransky2010occulting}. 

\subsubsection{Planet Physical Model}
The Planet Physical Model module contains models of the light emitted or reflected by planets in the wavelength bands under investigation by the current mission simulation.  It uses physical quantities sampled from the distributions defined in the Planet Population, including planetary mass, radius, and albedo, along with the physical parameters of the host star stored in the Target List module, to generate synthetic spectra or band photometry, as appropriate. The planet physical model is explicitly defined separately from the population statistics to enable studies of specific planet types under varying assumptions of orbital or physical parameter distributions, i.e., evaluating the science yield related to Earth-like planets under different definitions of the habitable zone.  The specific implementation of this module can vary greatly, and can be based on any of the many available planetary albedo, spectra and phase curve models\cite{Pollack1986,Marley1999,Fortney2008,Cahoy2010,Spiegel2012,burrows1997nongray,burrows2003beyond}. 

\subsubsection{Time}\label{sec:time}
The Time module is responsible for keeping track of the current mission time.  It encodes only the mission start time, the mission duration, and the current time within a simulation.  All functions in all modules requiring knowledge of the current time call functions or access parameters implemented within the Time module.  Internal encoding of time is implemented as the time from mission start (measured in days).  The Time module also provides functionality for converting between this time measure and standard measures such as Julian Day Number and UTC time.

\subsubsection{Rules}
The Rules module contains additional constraints placed on the mission design not contained in other modules. These constraints are passed into the Survey Simulation module to control the simulation. For example, a constraint in the Rules module could include prioritization of revisits to stars with detected exoplanets for characterization when possible. This rule would force the Survey Simulation module to simulate observations for target stars with detected exoplanets when the Observatory Module determines those stars are observable.

The Rules module also encodes the calculation of integration time for an observation.  This can be based on achieving a pre-determined signal to noise (SNR) metric (with various possible definitions), or via a probabilistic description as in \refnum{kasdin2006}.  This requires also defining a model for the background contribution due to all astronomical sources and especially due to zodiacal and exozodiacal light\cite{Stark2014}.

The integration time calculation can have significant effects on science yield---integrating to the same SNR on every target may represent a suboptimal use of mission time, as could integrating to achieve the minimum possible contrast on very dim targets.  Changing the implementation of the Rules module allows exploration of these tradeoffs directly.

\subsubsection{Post-Processing}
The Post-Processing module encodes the effects of post-processing on the data gathered in a simulated observation, and the effects on the final contrast of the simulation.  In the simplest implementation, the Post-Processing module does nothing and simply assumes that the attained contrast is some constant value below the instrument's designed contrast---that post-processing has the effect of uniformly removing background noise by a pre-determined factor.  A more complete implementation actually models the specific effects of a selected post-processing technique such as LOCI\cite{lafreniere2007new} or KLIP\cite{soummer2012detection} on both the background and planet signal via either processing of simulated images consistent with an observation's parameters, or by some statistical description.

The Post-Processing module is also responsible for determining whether a planet detection has occurred for a given observation, returning one of four possible states---true positive (real detection), false positive (false alarm), true negative (no detection when no planet is present) and false negative (missed detection).  These can be generated based solely on statistical modeling as in \refnum{kasdin2006}, or can again be generated by actually processing simulated images.

\subsection{Simulation Modules}
The simulation modules include Target List, Simulated Universe, Survey Simulation and Survey Ensemble. These modules perform tasks which require inputs from one or more input modules as well as calling function implementations in other simulation modules.

\subsubsection{Target List}
The Target List module takes in information from the Optical System Description, Star Catalog, Planet Population Description, and Observatory Definition input modules and generates the input target list for the simulated survey.  This list can either contain all of the targets where a planet with specified parameter ranges could be observed\cite{savransky2008}, or can contain a list of pre-determined targets such as in the case of a mission which only seeks to observe stars where planets are known to exist from previous surveys.  The final target list encodes all of the same information as is provided by the Star Catalog module.

\subsubsection{Simulated Universe}
The Simulated Universe module takes as input the outputs of the Target List simulation module to create a synthetic universe composed of only those systems in the target list.  For each target, a planetary system is generated based on the statistics encoded in the Planet Population Description module, so that the overall planet occurrence and multiplicity rates are consistent with the provided distribution functions.  Physical parameters for each planet are similarly sampled from the input density functions.  This universe is encoded as a list where each entry corresponds to one element of the target list, and where the list entries are arrays of planet physical parameters.  In cases of empty planetary systems, the corresponding list entry contains a null array.

The Simulated Universe module also takes as input the Planetary Physical Model module instance, so that it can return the specific spectra due to every simulated planet at an arbitrary observation time throughout the mission simulation.

\subsubsection{Survey Simulation}
The Survey Simulation module takes as input the output of the Simulated Universe simulation module and the Time, Rules, and Post-Processing input modules. This is the module that performs a specific simulation based on all of the input parameters and models. This module returns the mission timeline - an ordered list of simulated observations of various targets on the target list along with their outcomes.  The output also includes an encoding of the final state of the simulated universe (so that a subsequent simulation can start from where a previous simulation left off) and the final state of the observatory definition (so that post-simulation analysis can determine the percentage of volatiles expended, and other engineering metrics).

\subsubsection{Survey Ensemble}
The Survey Ensemble module's only task is to run multiple simulations.  While the implementation of this module is not at all dependent on a particular mission design, it can vary to take advantage of available parallel-processing resources.  As the generation of a survey ensemble is an embarrassingly parallel task---every survey simulation is fully independent and can be run as a completely separate process---significant gains in execution time can be achieved with parallelization.  The baseline implementation of this module contains a simple looping function that executes the desired number of simulations sequentially, as well as a locally parallelized version based on IPython Parallel\cite{perez2007ipython}.

\section{WFIRST-AFTA Coronagraph Modeling}\label{sec:wfirst}
While the development of EXOSIMS is ongoing, we have already produced simulation results with the functionality out of which the baseline EXOSIMS implementation is being built.  In this section, we present the results of some mission simulations for WFIRST-AFTA using optical models of coronagraph designs generated at JPL during the coronagraph downselect process in 2013, as well as post-downselect optical models of the Hybrid Lyot Coronagraph (HLC)\cite{trauger2012complex} generated in 2014\footnote{J. Krist, personal communication, 2014}.  It is important to emphasize that the instrument designs and mission yields shown here are not representative of the final coronagraphic instrument or its projected performance.  All of the design specifics assumed in these simulations are still evolving in response to ongoing engineering modeling of the observatory as a whole and to best meet the mission science requirements.  

These simulations are instead presented in order to highlight the flexibility of the EXOSIMS approach to mission modeling, and to present two important use cases.  In \S\ref{sec:predown} we present mission yield comparisons for different instrument designs while all other variables (observatory, star catalog, planet models, etc.) are kept constant. The results from these simulations are most useful for direct comparisons between different instruments and to highlight particular strengths and weaknesses in specific designs. Ideally, they can be used to guide ongoing instrument development and improve the final design science yield. In \S\ref{sec:hlcparams} we investigate a single coronagraph design operating under varying assumptions on observatory stability and post-processing capabilities.  These simulations highlight how EXOSIMS can be used to evaluate a more mature instrument design to ensure good results under a variety of operating parameters.  This section also demonstrates how to incorporate the effects of different assumptions in the pre-simulation optical system diffractive modeling.

In addition to the HLC, the first set of optical models includes models for a Shaped Pupil Coronagraph (SPC)\cite{zimmerman2015shaped} and a Phase-Induced Amplitude Apodization Complex Mask Coronagraph (PIAA-CMC) \cite{sidick2014simulated}.  In the downselect process, the SPC and HLC were selected for further development with PIAA-CMC as backup.  It should be noted that the HLC optical models in the first and second set of simulations shown here represent different iterations on the coronagraph design, and thus represent different instruments.

The Optical System Description is implemented as a static point spread function, throughput curve, and contrast curve based on the JPL optical models.  Other values describing the detector, science instrument and the rest of the optical train were chosen to match \refnum{traub2014science} as closely as possible.  The integration times in the Rules module are determined via modified equations based on \refnum{kasdin2006} to achieve a specified false positive and negative rate, which are encoded as constant in the post-processing module. Spectral characterization times are based on pre-selected SNR values (as in \refnum{brown2005}) and match the calculations in \refnum{traub2014science}.  

The Star Catalog is based on a curated database originally developed by Margaret Turnbull\cite{turnbull2012search}, with updates to stellar data, where available, taken from current values from the SIMBAD Astronomical Database\cite{wenger2000simbad}.  Target selection is performed with a detection integration time cutoff of 30 days and a minimum completeness cutoff of 2.75\%\cite{savransky2008}. Revisits are permitted at the discretion of the automated scheduler\cite{Savransky2010}, and one full spectrum is attempted for each target (spectra are not repeated if the full band is captured on the first attempt).  The total integration time allotted is one year, spaced over six years of mission time with the coronagraph getting top priority on revisit observations.  

\subsection{Comparison of Pre-Downselect Coronagraph Designs}\label{sec:predown}

As a demonstration of EXOSIMS ability to compare different instrument designs for a single mission concept, we compare mission simulation results based on optical models of the pre-downselect SPC, HLC and PIAA-CMC designs.  As all of these represent preliminary designs that have since been significantly improved upon, and as our primary purpose here is to demonstrate the simulations' utility, we will refer to the three coronagraphs simply as C1, C2, and C3 (in no particular order). \reftable{tbl:corons} lists some of the parameters of the three coronagraphs including their inner and outer working angles, their minimum and mean contrasts, and maximum and mean throughputs.  Each design has significantly different operating characteristics in its region of high contrast (or `dark hole'). C3 provides the best overall minimum contrast and IWA, but has a more modest mean contrast, whereas C2 has the most stable, and lowest mean contrast over its entire dark hole, at the expense of a larger inner working angle.  C1 has the smallest angular extent for its dark hole, but maintains reasonably high throughput throughout.  C2 has a constant, and very low throughput, while C3 has the highest throughput over its entire operating region.  Finally, while C1 and C3 cover the full field of view with their dark holes, C2 only creates high contrast regions in 1/3 of the field of view, and so requires three integrations to cover the full field.

We consider five specific metrics for evaluating these coronagraph designs:
\begin{enumerate}
\item Unique planet detections, defined as the total number of individual planets observed at least once.
\item All detections, defined as the total number of planet observations throughout the mission (including repeat observations of the same planets).
\item Total visits, defined as the total number of observations.
\item Unique targets, defined as the number of target stars observed throughout the mission.
\item Full spectral characterizations, defined as the total number of spectral characterizations covering the entire 400 to 800 nm band. This does not include characterizations where the inner or outer working angle prevent full coverage of the whole band. This number will always be smaller than the number of unique detections based on the mission rules used here.
\end{enumerate}
While it is possible to use EXOSIMS results to calculate many other values, these metrics represent a very good indicator of overall mission performance.  As it is impossible to jointly maximize all five---in particular, getting more full spectra or additional repeat detections is a direct trade-off to finding additional, new planets---these values together describe the Pareto front of the mission phase space.  At the same time, these metrics serve as proxies for other quantities of interest.  For example, taken together, all detections and unique detections indicate a mission's ability to confirm it's own detections during the course of the primary mission, as well as for possible orbit fitting to detected planets.  The number of unique targets, compared with the input target list, determines whether a mission is operating in a `target-poor' or `execution time-poor' regime.  The latter can be addressed simply by increasing the mission lifetime, whereas the former can only be changed with an instrument redesign.   Finally, comparing the numbers of unique detections and full spectra indicates whether an instrument design has sufficient capabilities to fully characterize the planets that it can detect.

\begin{figure}[ht]
\centering
\includegraphics[width=0.65\textwidth]{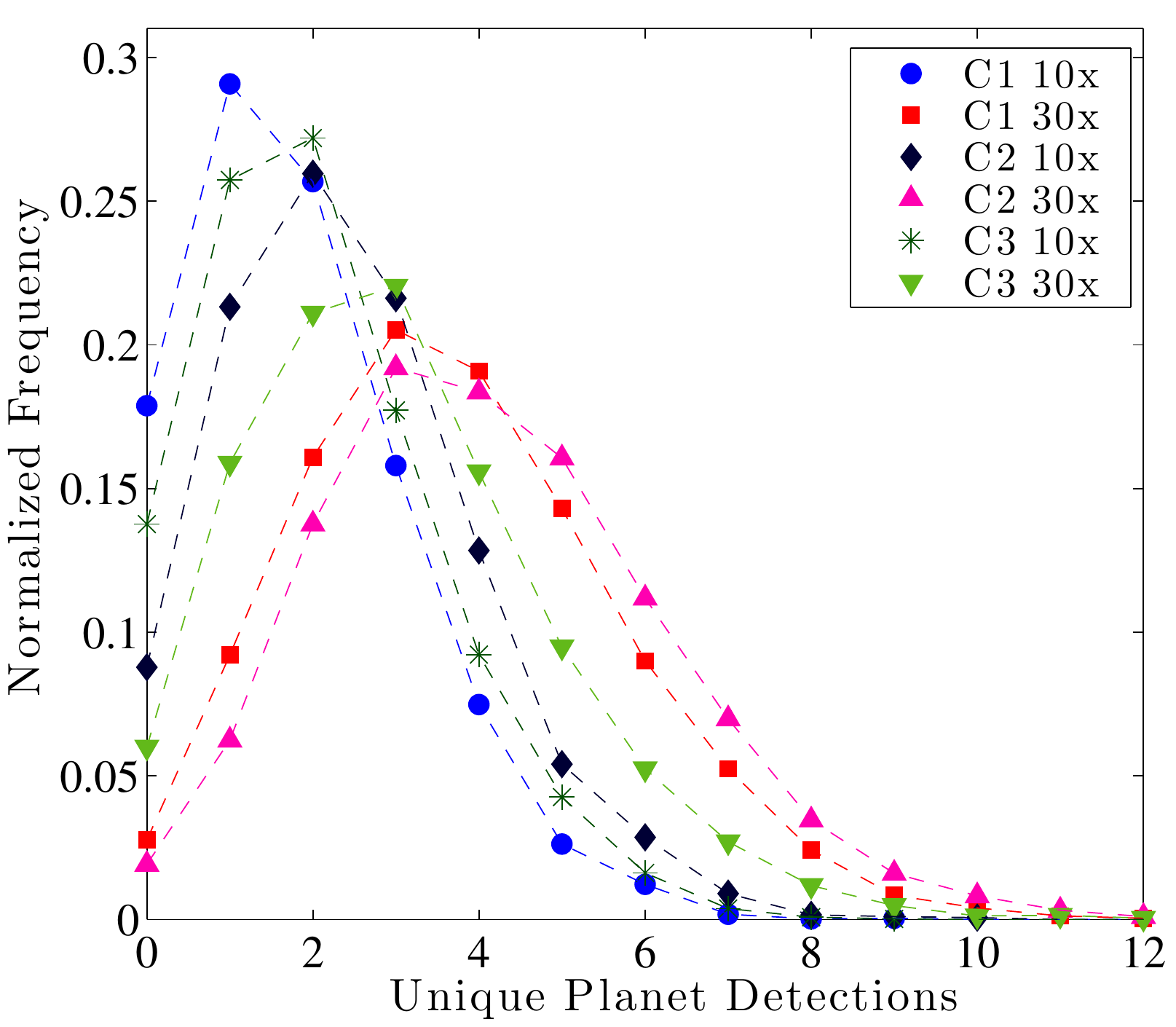}
\caption{PDF of unique detections (number of individual planets, potentially with multiple planets about some targets, detected one or more times) for the coronagraph designs described in \reftable{tbl:corons} assuming either a factor of 10 or 30 in post-processing contrast gains. Of particular importance here is the probability of zero detections---all of the designs at 10x suppression, and C1 in particular, have a significant ($>5\%$) chance of never seeing a planet. \label{fig:audets}}
\end{figure}

\begin{figure}[ht]
\centering
\includegraphics[width=0.65\textwidth]{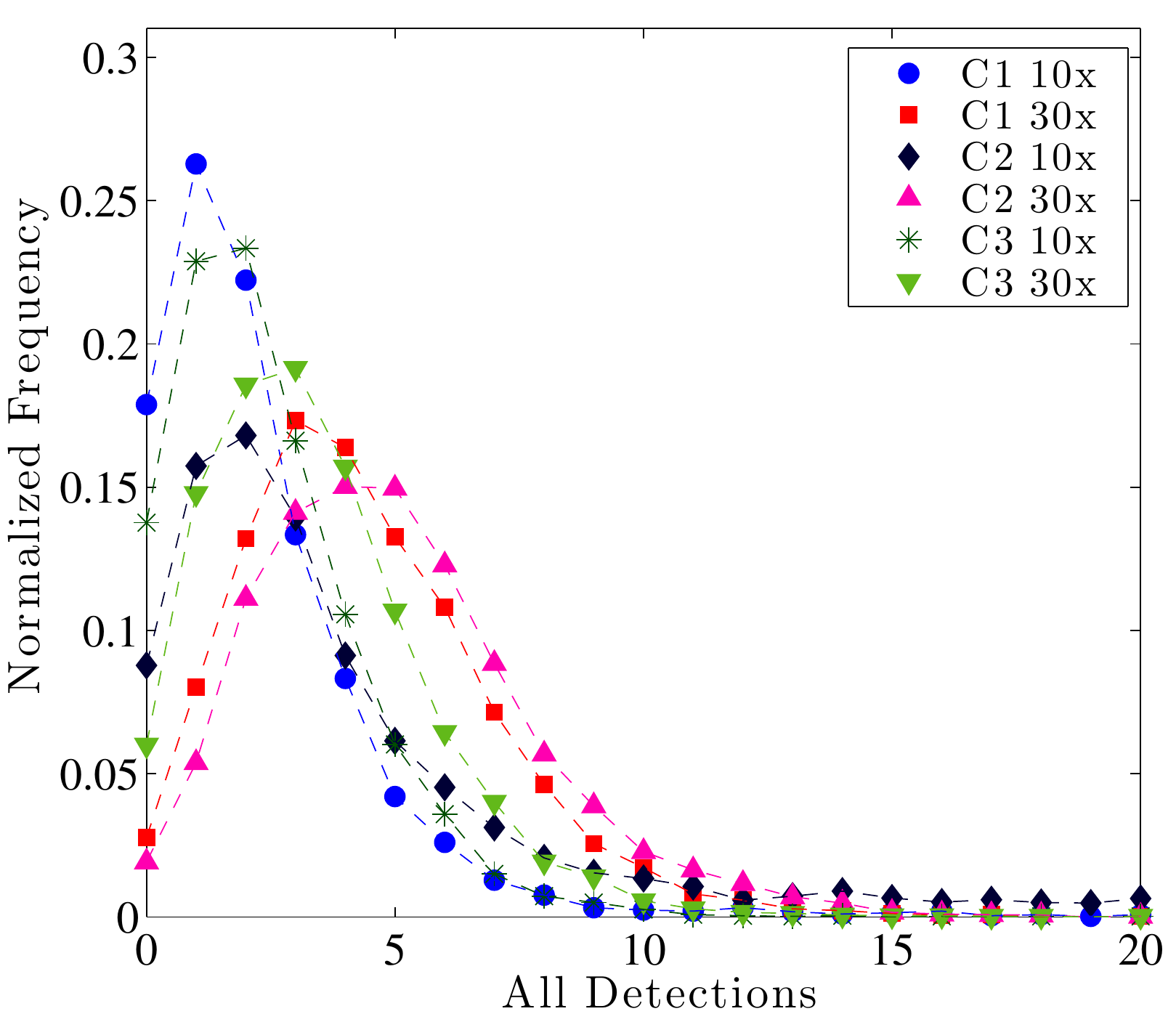}
\caption{PDF of all detections (including repeat detections) for instruments as in \reffig{fig:audets}. Note that values of 15 or more typically represent a small number of easily detectable planet that are re-observed many times.  Re-observations of a single target were capped at four successful detections in all simulations. \label{fig:adets}}
\end{figure}

\begin{figure}[ht]
\centering
\includegraphics[width=0.65\textwidth]{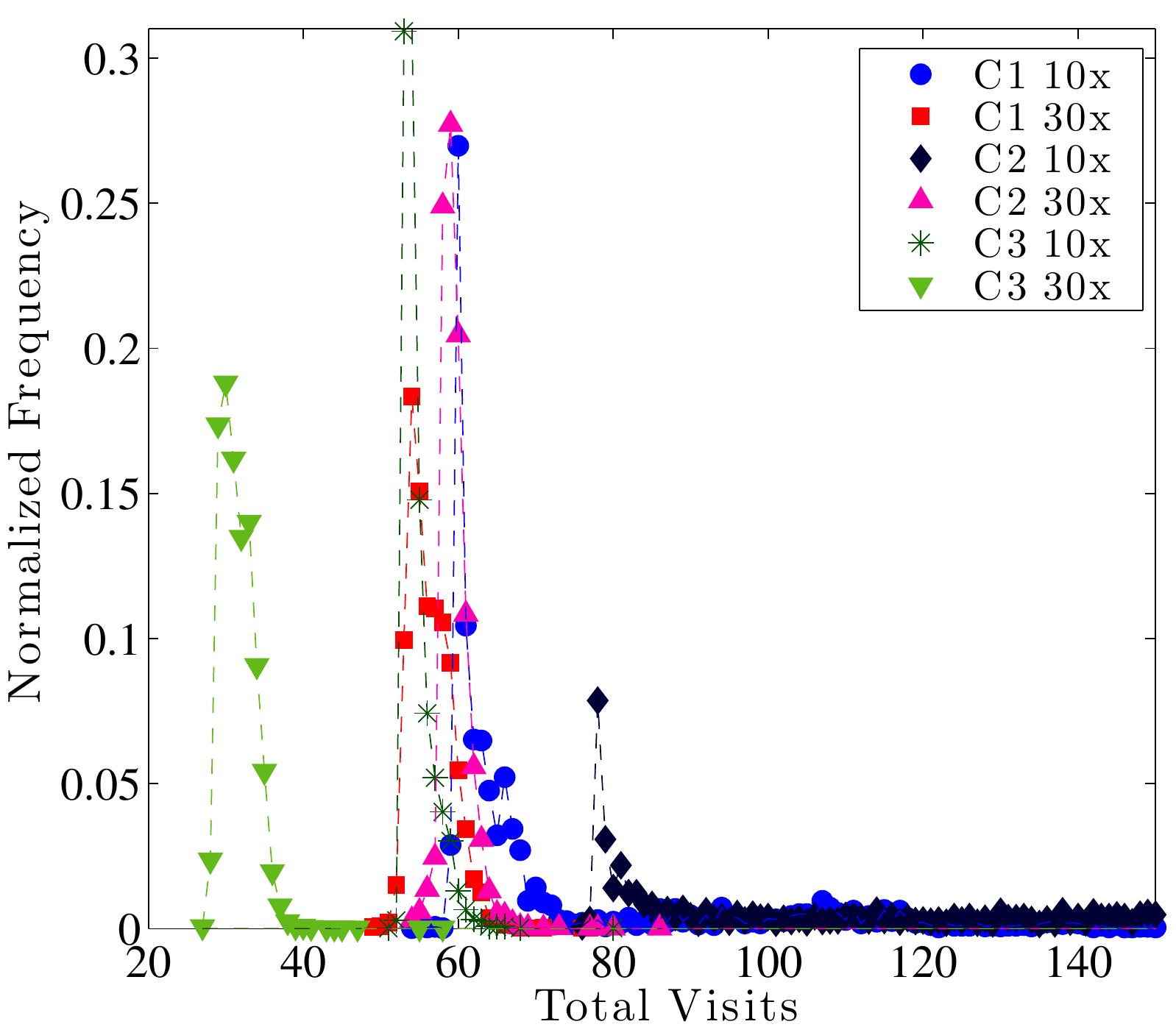}
\caption{PDF of total number of observations (including repeat observations of some targets) for instruments as in \reffig{fig:audets}.\label{fig:avisits}}
\end{figure}

\begin{figure}[ht]
\centering
\includegraphics[width=0.65\textwidth]{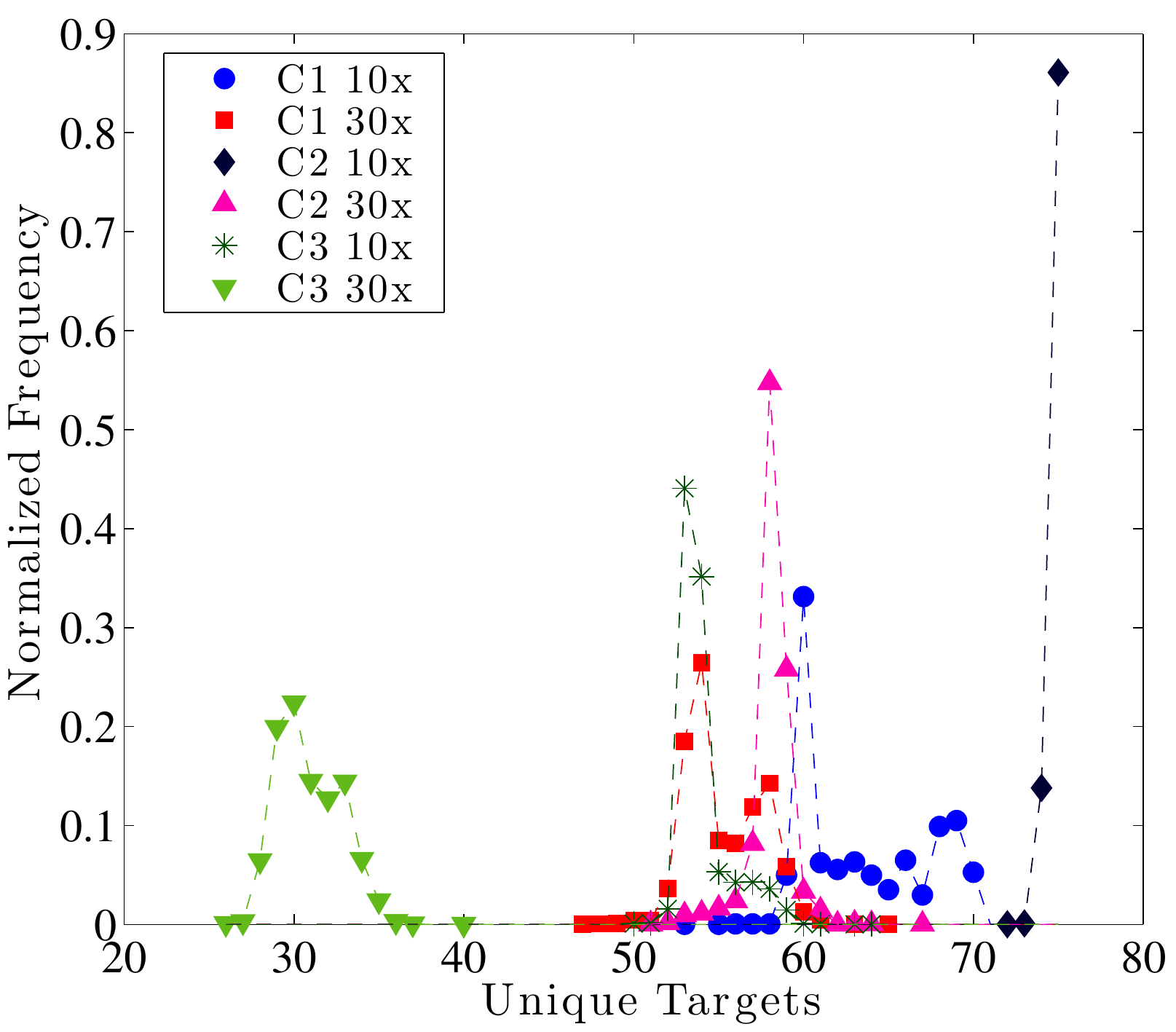}
\caption{PDF of unique targets observed for instruments as in \reffig{fig:audets}.  While all three instruments have fairly narrow distributions of this parameter, only C2 with 10x post-processing gains is completely target limited.\label{fig:auvisits}}
\end{figure}

\begin{figure}[ht]
\centering
\includegraphics[width=0.65\textwidth]{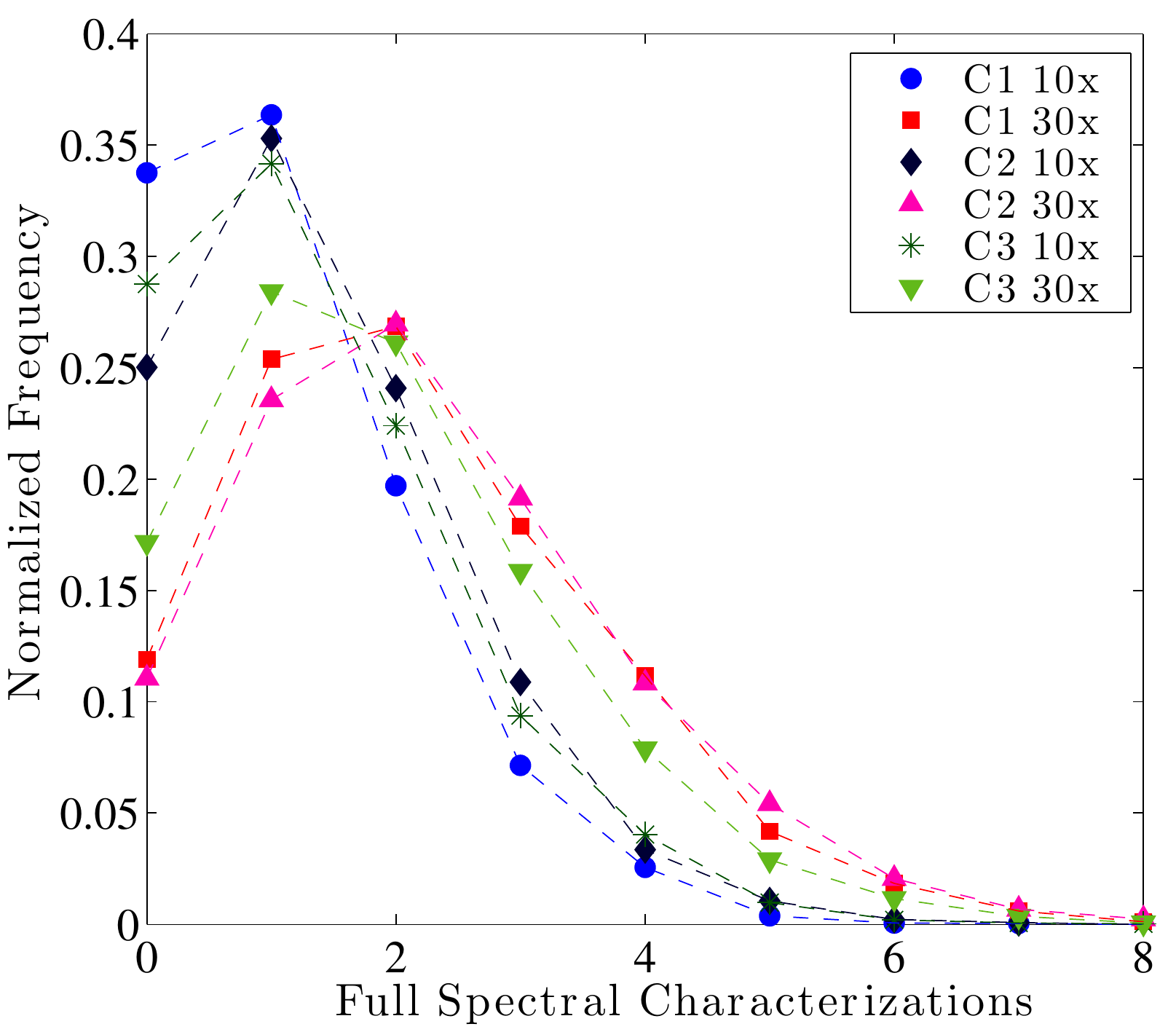}
\caption{PDF of number of spectra achieved over the whole band from 400 to 800 nm for instruments as in \reffig{fig:audets}. C3 does comparatively well in this metric due to its lower IWA and high throughput. \label{fig:spectra}}
\end{figure}

For each of the coronagraphs we run 5000 full mission simulations, keeping all modules except for the Optical Description and Post-Processing constant. In addition to the parameters and implementations listed above, our Post-Processing module implementation assumes a static factor of either 10 or 30 in terms of contrast improvement due to post-processing.  That is, results marked 10x assume that the achieved contrast on an observation is a factor of 10 below the design contrast at the equivalent angular separation. All together, we generated 30,000 discrete mission simulations, in six ensembles.  Mean values and $1\sigma$ standard deviations for our five metrics of interest for each ensemble are tabulated in \reftable{tbl:res2}, with the full probability density functions (PDFs) shown in Figs.~\ref{fig:audets} - \ref{fig:spectra}. 

From the tabulated values, we see that the three coronagraphs have fairly similar performance in terms of number of planets found and spectrally characterized.  Overall, C2 is most successful at detecting planets, due primarily to the stability of its contrast over the full dark hole.  However, because of the very low overall throughput, this does not translate into more spectral characterizations than the other two designs.  C1 and C2 benefit more from the change from 10x to 30x contrast improvement due to post-processing than does C3, which already has the deepest overall contrast, but whose contrast varies significantly over the dark hole.  The largest differences in the metrics are the total number of observations.  These illustrate the direct trade-off between acquiring spectra, which take a very long time, and doing additional integrations on other targets. In cases such as C2 with only 10x contrast improvement, the spectral characterization times are typically so long that most targets do not stay out of the observatory's keepouts and so the mission scheduling logic chooses to do more observations rather than wasting time on impossible spectral integrations. 

Turning to the figures of the full distributions for these metrics, we see that despite having similar mean values for unique planet detections, the full distributions of detections are quite different, leading to varying probabilities of zero detections.  As this represents a major mission failure mode, it is very important to track this value, as it may outweigh the benefits of a given design.  C1 with only 10x contrast gain does particularly poorly in this respect, with over 15\% of cases resulting in no planets found.  However, when a 30x gain is assumed, C1 and C2 end up having the lowest zero detection probabilities.  We again see that the effects of even this simple post-processing assumption are not uniform over all designs.  This is due to the complicated interactions between each instrument's contrast curve and the assumed distributions of planetary parameters.  In essence, if our priors were different (leading to different completeness values for our targets) then we would expect different relative gains for the same post-processing assumptions. This is always a pitfall of these simulations and must always be kept in mind when analyzing the results.  It should also be noted that there have been multiple iterations of all these coronagraph designs since downselect, resulting in significantly lower probabilities of zero detections, as seen in the next section.

Another interesting feature are the very long right-hand tails of the all detections and total visits distributions.  These do not actually represent outliers in terms of highly successful missions, but rather typically imply the existence of one or a small number of very easy to detect planets.  The logic of the scheduler allows the mission to keep returning to these targets for followup observations when it has failed to detect any other planets around the other targets in its list.  This situation arises when the design of the instrument and assumptions on planet distributions leave a mission target limited.  The distributions of unique targets show this limitation, with very narrow density functions for the actual number of targets observed for each instrument.  In particular, \reffig{fig:auvisits} makes it clear that C2 with 10x post-processing gains runs out of available targets.  In order to combat this, the scheduler code prevents revisits to a given target after four successful detections of a planet around it.  Finally, turning to \reffig{fig:spectra} we see that all three designs, regardless of post-processing assumptions have greater than 10\% probabilities of zero full spectral characterizations.  C1 with 10x post-processing gains fares most poorly with zero full spectra achieved in over one third of all cases.

\begin{figure}[ht]
\centering
\includegraphics[width=0.65\textwidth]{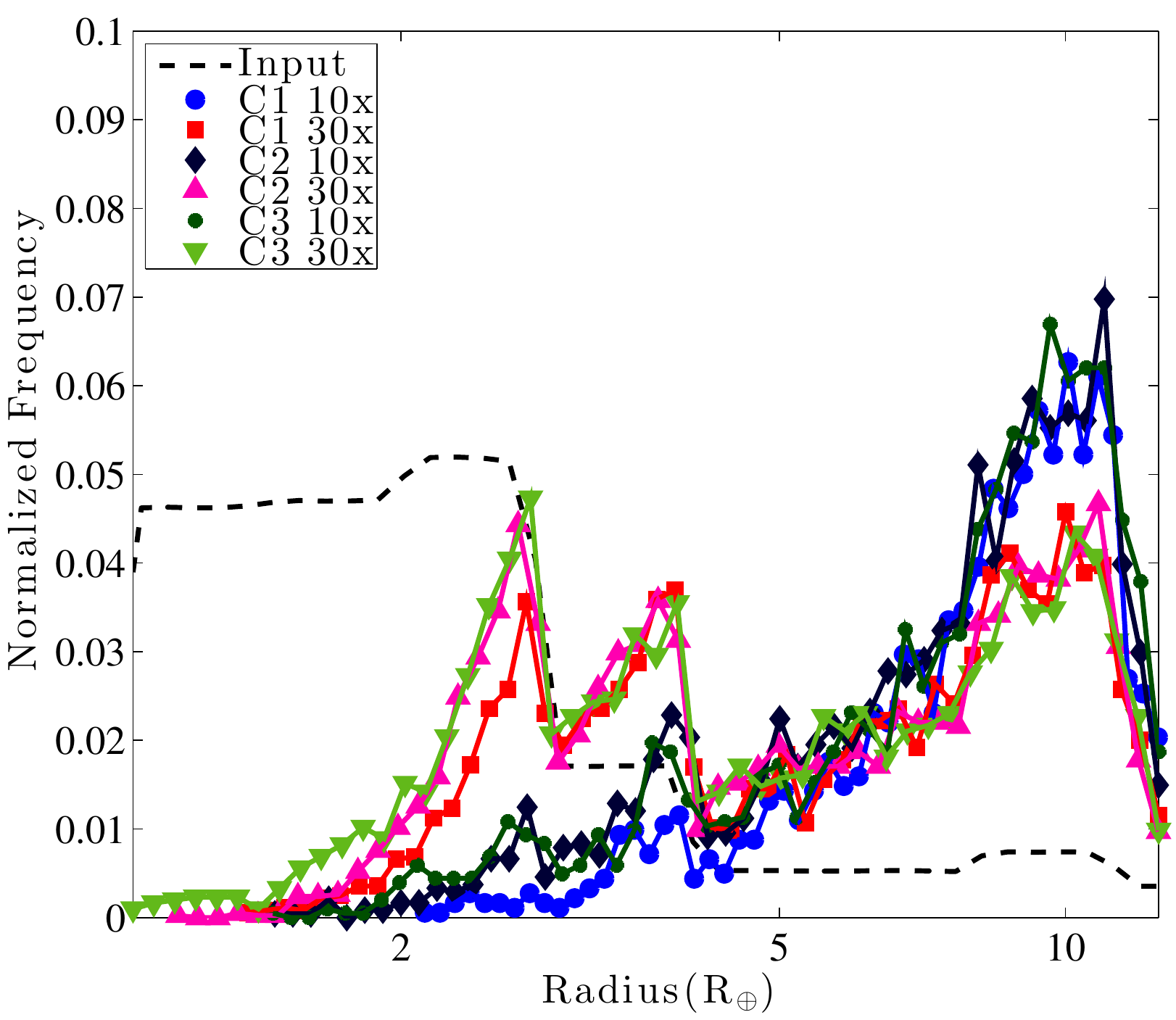}
\caption{Input and output distributions of planetary radius for instruments as in \reffig{fig:audets}.  The black dashed line represents the density function used in generating the planetary radii for the simulated planets in all simulations, while the other lines represent the distributions of planetary radii of the planets detected by each of the coronagraphs.  The input distribution is based on the Kepler results reported in \refnum{fressin2013false}. \label{fig:radiusdists}}
\end{figure}

\begin{figure}[ht]
\centering
\includegraphics[width=0.65\textwidth]{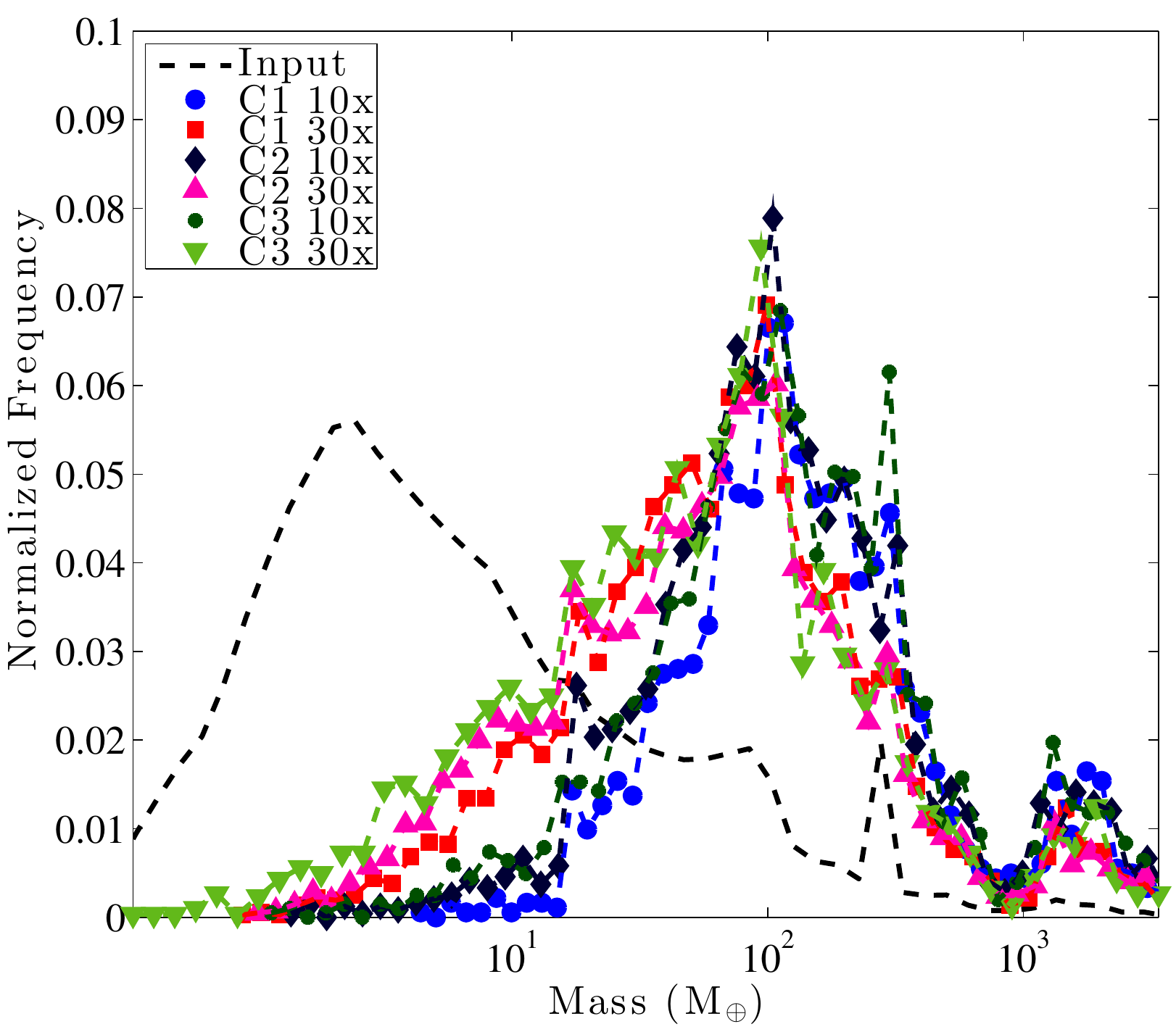}
\caption{Input and output distributions of planetary mass for instruments as in \reffig{fig:audets}.  The input mass distribution is derived from sampling the radius distribution shown in \reffig{fig:radiusdists} and converting to mass via an assumed density function. \label{fig:massdists}}
\end{figure}

Analysis of the survey ensembles also allows us to measure the biasing effects of the mission on the planet parameters of interest.  As we know the input distributions of the simulation, we can think of these as priors, and of the distribution of the `observed' planets as the posteriors.  Figs.~\ref{fig:radiusdists} and \ref{fig:massdists} show the distributions of planetary mass and radius used in the simulations, respectively, along with the output distributions from the various coronagraph designs.   The output distributions are calculated by taking the results of all of the simulations in each ensemble together, as the number of planets detected in each individual simulation is too small to produce an accurate distribution.  

The input mass distribution shown here is derived from the Kepler radius distribution as reported in \refnum{fressin2013false} and is calculated by assuming that this distribution is the same for all orbital periods and via an assumed density function\cite{Savransky2013}.  The frequency spike seen at around 20 Earth masses is due to a poor overlap in the density functions used in this part of the phase space.  This results in an equivalent spike in the posterior distributions, which slightly biases the results.  

All of the instruments have fairly similar selection biases, although C1 and C3, which have smaller inner working angles and higher throughputs, detect more lower mass/radius planets. The effects of the instruments are readily apparent in all cases: lower radius planets, which are predicted to occur more frequently than larger radius ones, are detected at much lower rates.

\subsection{Comparison of HLC Parameters}\label{sec:hlcparams}

In this section we present the results of survey ensemble analyses for a single instrument---a post-downselect HLC design---again assuming either 10x or 30x post-processing gains, and assuming either 0.4, 0.8, or 1.6 milliarcseconds of telescope jitter.  The jitter of the actual observatory will be a function of the final bus design and the operation of the reaction wheels, and its precise value is not yet known, which makes it important to evaluate how different levels of jitter may effect the achieved contrast and overall science yield.  The jitter is built directly into the optical system model encoded in the Optical System Description module (see Krist et al., this volume, for details), while the post-processing is treated as in the previous section.

\begin{figure}[ht]
\centering
\includegraphics[width=0.65\textwidth]{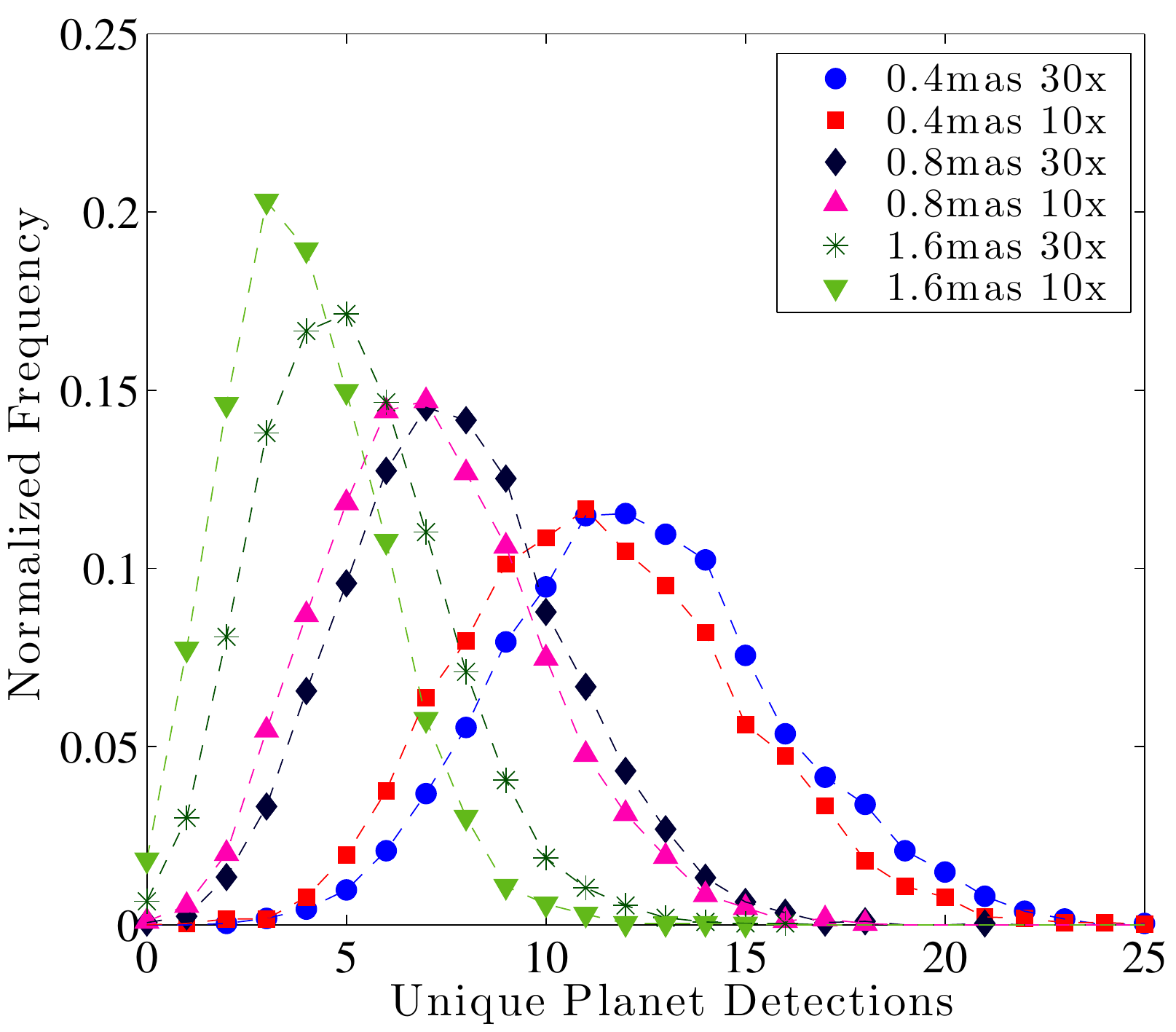}
\caption{PDF of unique planetary detections (number of individual planets, potentially with multiple planets about some targets, detected one or more times) for the post-downselect  HLC design, assuming either a factor of 10 or 30 in post-processing contrast gains and telescope jitter of 0.4, 0.8 or 1.6 mas. It should be noted that the change in assumed post-processing gain has a significantly smaller effect than the increased telescope jitter.  We also note that the 1.6 mas jitter cases still have a small (0.6 to 1.8\%) probability of never seeing a planet, whereas the 0.4 mas jitter ensembles do not contain a single simulation with zero plaents detected. \label{fig:audets2}}
\end{figure}

\begin{figure}[ht]
\centering
\includegraphics[width=0.65\textwidth]{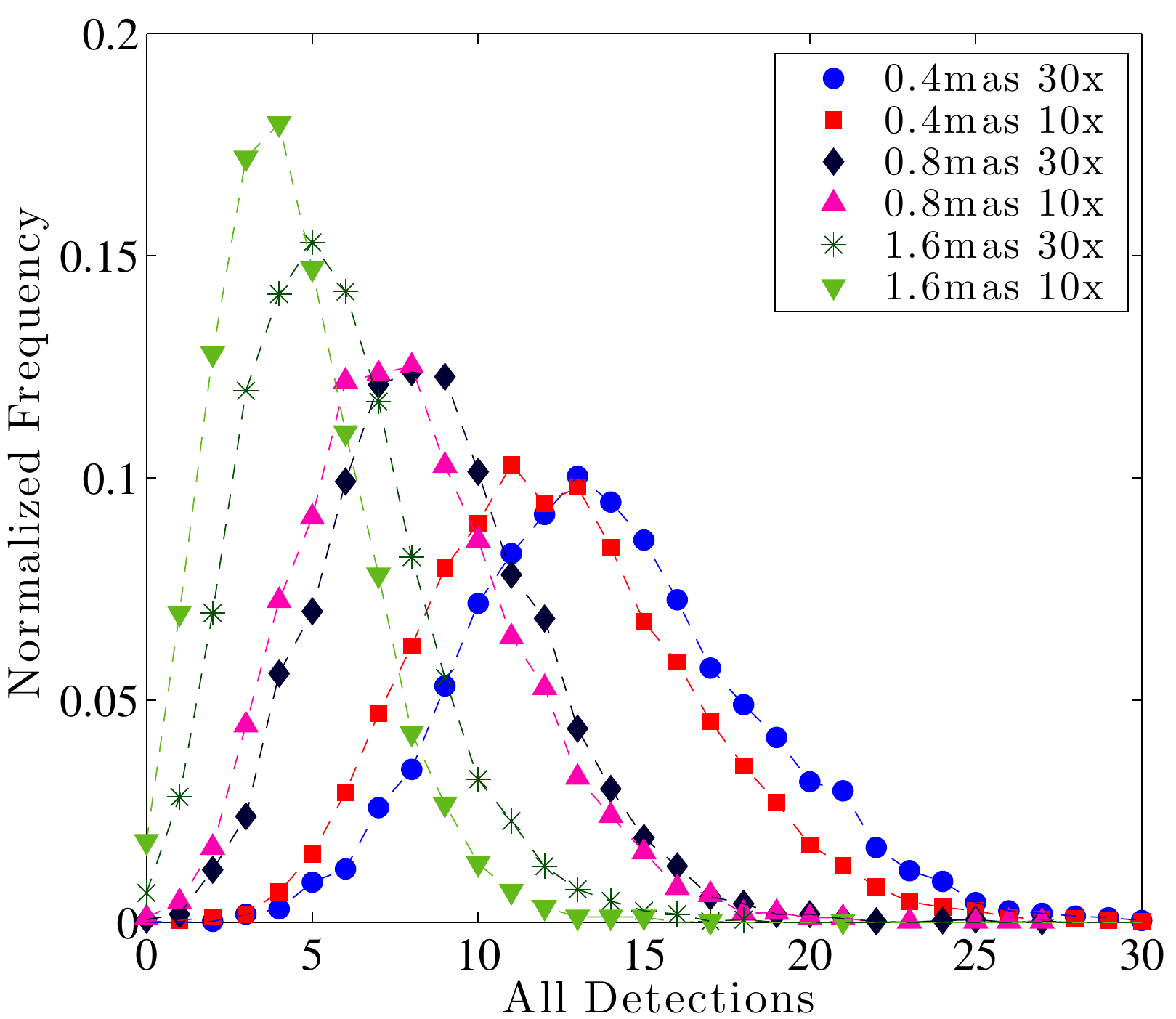}
\caption{PDF of total number of planetary detections (including repeat detections) for instruments as in \reffig{fig:audets2}. The trend here closely follows the one observed in the results for the unique detections metric.\label{fig:adets2}}
\end{figure}

\begin{figure}[ht]
\centering
\includegraphics[width=0.65\textwidth]{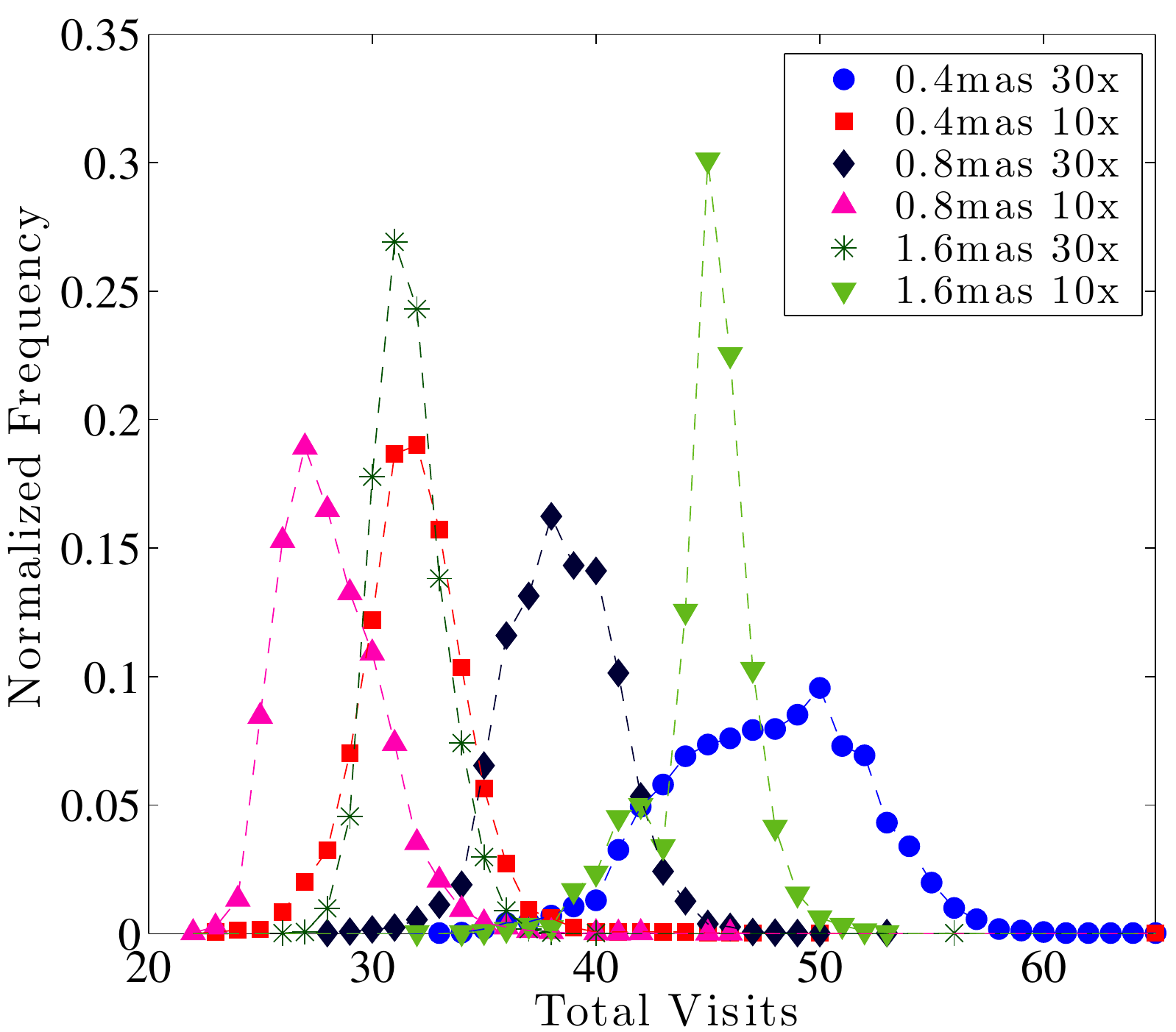}
\caption{PDF of total number of target observations (including repeat observations) for instruments as in \reffig{fig:audets2}. Here, the post-processing improvement factor makes more of a difference than in the previous two figures, as more time must be devoted to spectral characterizations, limiting how much time is available for further observations.\label{fig:auvisits2}}
\end{figure}

\begin{figure}[ht]
\centering
\includegraphics[width=0.65\textwidth]{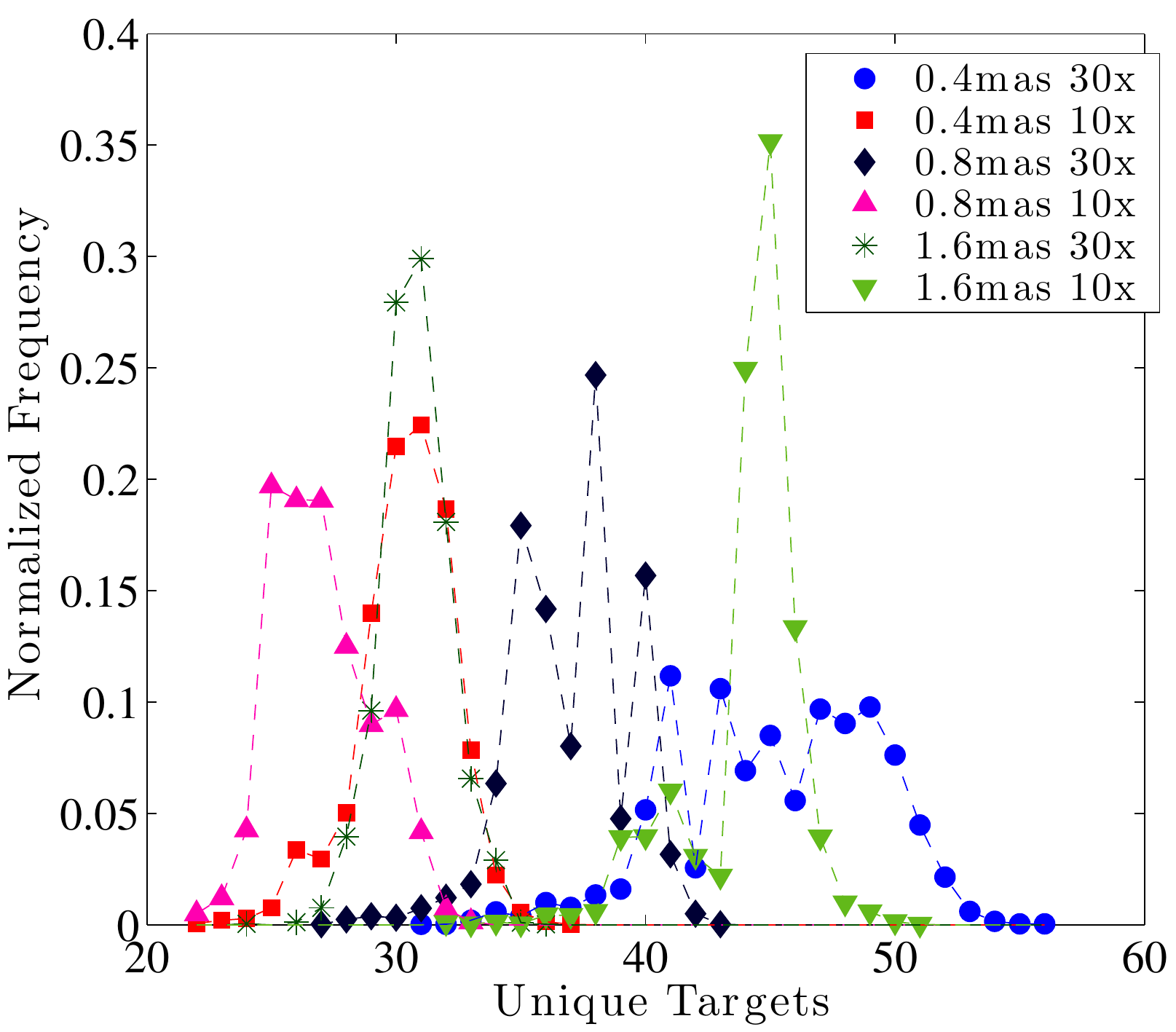}
\caption{PDF of unique targets observed for instruments as in \reffig{fig:audets2}.  The trend here tracks closely to the one observed in the total visits metric, and shows that this coronagraph design is not target limited in any of the studied cases.\label{fig:avisits2}}
\end{figure}

\begin{figure}[ht]
\centering
\includegraphics[width=0.65\textwidth]{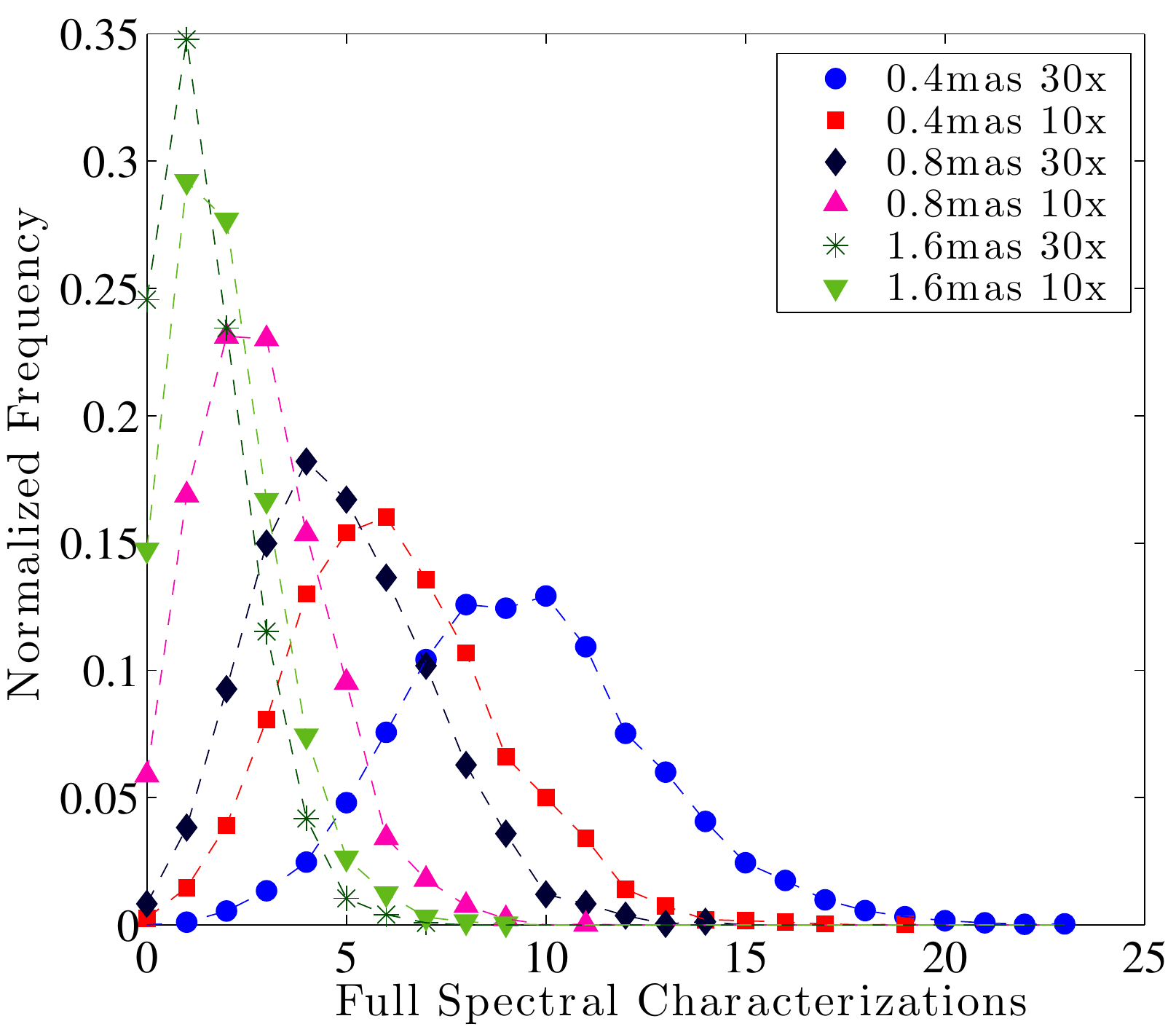}
\caption{PDF of number of spectra achieved over the whole band from 400 to 800 nm for instruments as in \reffig{fig:audets2}. In the worst case, there is an $\sim15\%$ of not getting any spectra.  Only the case of 0.4 mas jitter with 30x post-processing gain has no simulations in its ensemble with zero full spectra achieved.  \label{fig:spectra2}}
\end{figure}

As in the previous section, we run ensembles of 5000 simulations for each  of the six cases considered, keeping all modules except for the Optical Description and Post-Processing constant.  The mean and $1\sigma$ of the five metrics of interest described in \S\ref{sec:predown} are tabulated in \reftable{tbl:res3}, and the full PDFs for all metrics are shown in Figs.~\ref{fig:audets2} - \ref{fig:spectra2}.

One important observation made immediately obvious by these results is the relatively large effect of increased jitter versus the gains due to post-processing.  Tripling the assumed gain factor of post-processing on the final achieved contrast has a significantly smaller effect on the number of detections, gaining only one unique detection, on average, as compared with halving the amount of telescope jitter, which increases the number of unique detections by over 30\%, on average. This shows us that the telescope jitter may be an effect that fundamentally cannot be corrected after the fact, and therefore needs to be tightly controlled, with well defined requirements set during mission design.  Much of the current development effort for the project is focused on low-order wavefront sensing and control to mitigate these effects \cite{poberezhskiy2014technology,shi2015low}.

We can also see significant improvements in the coronagraph design since the versions evaluated in \S\ref{sec:predown}, as the probability of zero planet detections is less than 2\% in the case of the highest jitter level, and is well below 1\% for all other cases.  In fact, for both the 0.4 mas jitter ensembles, no simulations had zero detections, indicating a very low probability of complete mission failure for this coronagraph at these operating conditions.

Similar to the results of the previous section, the trend in the number of total visits does not simply follow those seen in the unique and total detection metrics, but is a function of both the number of detections and how much time is spent on spectral characterizations.  We can see how the cases with the highest jitter and lowest post-processing gains are pushed towards larger numbers of observations, and unique targets, as they are able to achieve fewer full spectral characterizations, leaving them with additional mission time to search for new candidates. This is equally reflected in \reffig{fig:spectra2} where, despite the good performance seen in \reffig{fig:audets2}, all jitter levels have over 5\% chance of zero full spectra at the 10x post-processing gain level, and only the 0.4 mas case at 30x gain has no instances of zero full spectra in its ensemble of results.

These metrics, taken together, clearly show that further optimization is possible via modification of mission rules, which were kept constant in all these ensembles.  For example, the low numbers of spectral characterizations at higher jitter levels suggest that it may be worthwhile to attempt shallower integrations in order to be able to make more total observations and potentially find a larger number of bright planets.  This would bias the final survey results towards larger planets, but would increase the probability of spectrally characterizing at least some of the planets discovered.  Alternatively, this may point to the desirability of investigating whether full spectral characterizations can be achieved for a small number of targets over the course of multiple independent observations.

\section{Conclusions}

We have presented the design details of EXOSIMS---a modular, open source software framework for the simulation of exoplanet imaging missions with instrumentation on space observatories.  We have also motivated the development and baseline implementation of the component parts of this software for the WFIRST-AFTA coronagraph, and presented initial results of mission simulations for various iterations of the WFIRST-AFTA coronagraph design.

These simulations allow us to compare completely different instruments in the form of early competing coronagraph designs for WFIRST-AFTA.  The same tools also allow us to evaluate the effects of different operating assumptions, demonstrated here by comparing different assumed post-processing capabilities and telescope stability values for a single coronagraph design.

As both the tools and the coronagraph and mission design continue to mature we expect the predictions presented here to evolve as well, but certain trends have emerged that we expect to persist.  We have identified the portions of design space and telescope stability ranges that lead to significant probabilities of zero detections, and we expect instrument designs and observatory specifications to move away from these.  We have also identified a mean number of new planetary detections, for our particular assumed prior distributions of planetary parameters, that are consistent with the science definition team's mission goals for this instrument.

As we continue to both develop the software and to improve our specific modeling of WFIRST-AFTA we expect that these and future simulations will prove helpful in guiding the final form of the mission, and will lay the groundwork for analysis of future exoplanet imagers. 

\acknowledgments 
This material is based upon work supported by the National Aeronautics and Space Administration under Grant No. NNX14AD99G issued through the Goddard Space Flight Center.  EXOSIMS is being developed at Cornell University with support by NASA Grant No. NNX15AJ67G.  This research has made use of the SIMBAD database,
operated at CDS, Strasbourg, France.  The authors would like to thank Rhonda Morgan for many useful discussions and suggestions, as well as our reviewers Wes Traub and Laurent Pueyo, who have significantly improved this work through their comments. 



\vspace{2ex}\noindent{\bf Dmitry Savransky} is an assistant professor in the Sibley School of Mechanical and Aerospace Engineering at Cornell University. He received his PhD from Princeton University in 2011 followed by a postdoctoral position at Lawrence Livermore National Laboratory where he assisted in the integration and commissioning of the Gemini Planet Imager.  His research interests include optimal control of optical system, simulation of space missions, and image post-processing techniques.

\vspace{2ex}\noindent{\bf Daniel Garrett} is a PhD student in the Sibley School of Mechanical and Aerospace Engineering at Cornell University.  His research interests include dynamics and control theory, planetary science, and space exploration.

\begin{table}[ht]
\caption{Parameters for coronagraphs studied in \S\ref{sec:predown}. \label{tbl:corons}} 
\begin{center}
\begin{tabular}{l c c c c c c c c}
     &     &     & \multicolumn{2}{c}{Contrast} & \multicolumn{2}{c}{Throughput$^b$} &   & FOV\\
Name & IWA$^a$ & OWA$^a$ & Min & Mean                   & Max & Mean                     & Sharpness$^c$ & Portion$^d$\\
\hline
C1 &  0.128 & 0.652 & 6.91e-09 & 1.60e-08 & 0.40 & 0.32 & 0.0142 & 1\\
C2 &  0.184 & 1.064 & 4.06e-09 & 7.06e-09 & 0.22 & 0.22 & 0.0138 & 1/3\\
C3 &  0.085 & 0.624 & 2.87e-09 & 2.94e-08 & 1.00 & 0.85 & 0.0143 & 1
\end{tabular}
\end{center}
\raggedright
\footnotesize $^a$Inner and outer working angle in arcseconds at 550 nm.\\
$^b$This is the throughput due to the coronagraph optics only. \\
$^c$Sharpness is defined as $\left(\sum_{i} P_i^2\right)/\left(\sum_i P_i\right)^2$ for normalized PSF $P_i$.\\
$^d$The fraction of the field of view covered by the coronagraph's region of high contrast.\\
\normalsize
\end{table}%

\begin{table}[ht]
\caption{Mean values and standard deviations of five performance metrics calculated from ensembles of mission simulations for the instruments described in \reftable{tbl:corons}.\label{tbl:res2}} 
\begin{center}
\begin{tabular}{l c | c c | c c | c c | c c | c c}
           &  &  \multicolumn{2}{c}{Unique}   &  \multicolumn{2}{c}{All}  & \multicolumn{2}{c}{Full}   &  \multicolumn{2}{c}{All}  &  \multicolumn{2}{c}{Unique} \\
 &Contrast & \multicolumn{2}{c}{Detections$^b$} & \multicolumn{2}{c}{Detections$^c$} & \multicolumn{2}{c}{Spectra$^d$} & \multicolumn{2}{c}{Visits} & \multicolumn{2}{c}{Targets}\\
Name & Factor$^a$ & $\mu$ & $1\sigma$ & $\mu$ & $1\sigma$ & $\mu$ & $1\sigma$ & $\mu$ & $1\sigma$ & $\mu$ & $1\sigma$\\ 
\hline
\multirow{2}{*}{C1} & 10x  &  1.8 &  1.4  &  2.6 &  4.4  &  1.1 &  1.1 &  74.9 & 28.2 &  63.5 &  3.7\\
 & 30x  &  3.7 &  2.0  &  4.4 &  2.6  &  2.2 &  1.5 &  56.4 & 2.7 &  55.3 &  2.2\\
 \hline
\multirow{2}{*}{C2} & 10x  &  2.4 &  1.6  &  7.8 &  13.5  &  1.4 &  1.2 &  141.3 & 38.8 &  74.9 &  0.4\\
& 30x  &  4.2 &  2.1  &  5.0 &  2.9  &  2.2 &  1.5 &  59.5 & 1.9 &  58.1 &  1.2\\
\hline
\multirow{2}{*}{C3} & 10x  &  2.0 &  1.5  &  2.4 &  2.0  &  1.3 &  1.2 &  54.7 & 2.0 &  54.0 &  1.5\\
 & 30x  &  3.0 &  1.9  &  3.4 &  2.3  &  1.9 &  1.4 &  31.5 & 2.2 &  30.9 &  1.9\\
\end{tabular}
\end{center}
\raggedright
\footnotesize $^a$Contrast improvement factor due to post-processing.\\
$^b$Number of individual planets detected one or more times.  \\
$^c$Total number of detections (including repeat detections of the same planets).\\
$^d$Total number of planets where spectra can be obtained over the whole wavelength range (400-800 nm).\\
\normalsize
\end{table}%

\begin{table}[ht]
\caption{Mean values and standard deviations of five performance metrics calculated from ensembles of mission simulations for the post-downselect HLC with varying levels of assumed telescope jitter.  Column definitions as in \reftable{tbl:res2}.\label{tbl:res3}} 
\begin{center}
\begin{tabular}{l c | c c | c c | c c | c c | c c}
           &  &  \multicolumn{2}{c}{Unique}   &  \multicolumn{2}{c}{All}  & \multicolumn{2}{c}{Full}   &  \multicolumn{2}{c}{All}  &  \multicolumn{2}{c}{Unique} \\
Jitter &Contrast & \multicolumn{2}{c}{Detections$^b$} & \multicolumn{2}{c}{Detections$^c$} & \multicolumn{2}{c}{Spectra$^d$} & \multicolumn{2}{c}{Visits} & \multicolumn{2}{c}{Targets}\\
(mas) & Factor$^a$ & $\mu$ & $1\sigma$ & $\mu$ & $1\sigma$ & $\mu$ & $1\sigma$ & $\mu$ & $1\sigma$ & $\mu$ & $1\sigma$\\ 
\hline
\multirow{2}{*}{0.4} & 30x  &  12.4 &  3.5  &  14.0 &  4.4  &  9.5 &  3.2 &  47.6 & 4.3 &  45.2 &  4.0\\
 & 10x  &  11.4 &  3.5  &  12.5 &  4.2  &  6.2 &  2.6 &  31.9 & 2.4 &  30.4 &  1.9\\
\multirow{2}{*}{0.8}  & 30x  &  7.8 &  2.8  &  8.7 &  3.3  &  4.9 &  2.2 &  38.4 & 2.5 &  37.0 &  2.3\\
 & 10x  &  7.2 &  2.7  &  8.0 &  3.3  &  2.8 &  1.7 &  28.1 & 2.3 &  27.0 &  2.0\\
\multirow{2}{*}{1.6} & 30x  &  5.1 &  2.3  &  5.7 &  2.7  &  1.4 &  1.2 &  31.6 & 1.6 &  30.8 &  1.4\\
& 10x  &  4.0 &  2.0  &  4.4 &  2.4  &  1.9 &  1.4 &  44.9 & 2.2 &  44.1 &  2.2\\
\end{tabular}
\end{center}
\end{table}%

\listoffigures

\end{spacing}
\end{document}